\documentstyle{article}
\title{Legendre transformations on the triangular lattice\footnotetext{
       The work is supported by grant RFBR-96-01-00128}}
\author{V.E. Adler\\ \\ \small \it
 Ufa Institute of Mathematics, Chernyshevsky str. 112, 450000 Ufa, Russia\\
 \small \it e-mail: adler@imat.rb.ru}
\date{17 August 1998}

\font\Sets=msbm10  \def\Real{\hbox{\Sets R}}
\def\phi{\varphi}  \def\eps{\varepsilon}
\def\12{{1\over2}} \def\LRA{\ \Leftrightarrow\ }
\def\la{\lambda}   \def\const{\mathop{\rm const}\nolimits}
\def\pa{\partial}  \def\ti{\widetilde}
\def\be  {\begin{equation}}  \def\ee  {\end{equation}}
\def\ba  {\begin{array}}     \def\ea  {\end{array}}
\def\bea {\begin{eqnarray}}  \def\eea {\end{eqnarray}}
\def\bean{\begin{eqnarray*}} \def\eean{\end{eqnarray*}}
\newtheorem{theorem}{Theorem}
\newtheorem{lemma}  {Lemma}
\newtheorem{Def}    {Definition}
\def\proof{\paragraph{Proof.}}
\def\qed{\vrule height0.6em width0.3em depth0pt}

\begin{document}
\maketitle
\thispagestyle{empty}

\section{Dual systems}

Study of the nonlinear lattices
\be                                                         \label{TL}
    \ddot q_n = r(\dot q_n)(g_{n+1}-g_n), \quad g_n=g(q_n-q_{n-1})
\ee
have brought Toda to the notion of dual system, which is obtained from the
original one by "replacement of particles by springs and springs by particles
accordingly to the certain rules" \cite{Toda}.  These rules are the
following.  The lattice (\ref{TL}) is equivalent to the Euler-Lagrange
equations for the functional
\[
    {\cal L}= \int dt\sum_n (a(\dot q_n)-b_n), \quad a''=1/r, \quad b'=g,
\]
where $a$ and $b$ define kinetic and potential energies respectively.  The
momentum conservation law ($T_n$ denotes shift operator $q_n\to q_{n+1}$)
\[
    {d\over dt}a'(\dot q_n) = (T_n-1)g_n
\]
allows to introduce the new variable $Q_n$ accordingly to the formulae
\[
    \dot Q_n= g_n, \quad Q_{n+1}-Q_n= a'(\dot q_n).
\]
Following the terminology of the paper \cite{AS}, we call this and analogous
changes generalized Legendre transformations.  The variable $Q_n$ satisfies
the dual lattice
\[
    \ddot Q_n = R(\dot Q_n)(G_{n+1}-G_n)
\]
where the functions $R$ and $G$ are defined by equations $g'=R(g),$
$G(a'(q))=q.$  For example, the Toda lattice $\ddot
q_n=\exp(q_{n+1}-q_n)-\exp(q_n-q_{n-1})$ is dual to the lattice $\ddot
Q_n=\dot Q_n(Q_{n+1}-2Q_n-Q_{n-1}).$

The following features of the considered transformation should be mentioned.
Firstly, it is possible due to the invariance of the Lagrangian with respect
to the shift $q\to q+\const.$  Secondly, it can be performed for arbitrary
$r$ and $g,$ and therefore it is not related with integrability.  At last, it
is involutory (up to the shift $T_n$) and therefore it cannot be used for
reproducing of solutions of the lattice.

It was shown in \cite{AS} that the situation changes drastically in the case
of the relativistic Toda type lattices
\be                                                         \label{RTL}
  \ddot q_n= r(\dot q_n)(\dot q_{n+1}h_{n+1}-\dot q_{n-1}h_n+g_{n+1}-g_n).
\ee
The Lagrangian for the lattice (\ref{RTL}) is obtained from the one written
above by adding of the term $-\dot q_nc_n,$ $c'=h$ which is interpreted as
the presence of magnetic field.  This term is equivalent, up to the total
divergence, to $-\dot q_nc_{n+1},$ and this leads to two equivalent forms of
the momentum conservation law
\[
    {d\over dt}(a'(\dot q_n)-c_n) =
       (T_n-1)(\dot q_nh_n+g_n) \LRA
\]\[
    \LRA  {d\over dt}(a'(\dot q_n)-c_{n+1}) =
              (T_n-1)(\dot q_{n-1}h_n+g_n)
\]
and consequently to the pair of generalized Legendre transformations
\bean
    T_-:&& \dot Q_n = \dot q_nh_n+g_n,    \quad
           Q_{n+1}-Q_n = a'(\dot q_n)-c_n,            \\
    T_+:&& \dot{\ti Q_n} = \dot q_{n-1}h_n+g_n,   \quad
           \ti Q_{n+1}-\ti Q_n = a'(\dot q_n)-c_{n+1}.
\eean
In contrast to the previous example, these transformations generally lead out
off the class (\ref{RTL}).  The requirement that the transformed lattice must
be of the same form turned out to be severe enough to isolate exactly the
subclass of the integrable lattices (that is, possessing higher symmetries
and conservation laws) \cite{AS}.  It is not so surprisingly, because in this
case the combination $T^{-1}_-T_+$ of the Legendre transformations defines
the B\"acklund autotransformation of the lattice under consideration, and
this property is indispensable feature of any integrable system.  The
situation is quite analogous to the case of the B\"acklund transformation for
the KdV equation which is composition of two slightly different Miura
transformations.

\begin{figure}[t]
\setlength{\unitlength}{0.07em} \small
\begin{picture}(290,240)(-230,-100)
 \put(-115,0){\vector(1,0){260}}
 \put(0,-115){\vector(0,1){260}}
 \put(-100,-100){\line(1,1){200}}
 \multiput(-100,-100)(200,100){2}{\line(0,1){100}}
 \multiput(-100,-100)(100,200){2}{\line(1,0){100}}
 \multiput(-100,0)(100,-100){2}{\line(1,1){100}}
 \put(150,-20){$T_m$} \put(-25,140){$T_n$}
 \put(-60,  5){$x$}        \put( 40,  5){$x_{1,0}$}
 \put(-60,-95){$x_{0,-1}$} \put( 40,105){$x_{1,1}$}
 \put(-95,-40){$y_{-1,0}$} \put(  5,-40){$y$}
 \put(  5, 60){$y_{0,1}$}  \put(105, 60){$y_{1,1}$}
 \put(-45,-55){$z$}        \put( 55,-55){$z_{1,0}$}
 \put(-45, 45){$z_{0,1}$}  \put( 55, 45){$z_{1,1}$}
\end{picture}\normalsize
\end{figure}
The aim of the present paper is to apply these ideas in the totally discrete
case: the differentiation $d/dt$ is replaced by the shift $T_m:q_{m,n}\to
q_{m+1,n}$ on the second discrete variable.  In order to abridge notation we
will omit, as a rule, the subscripts $m,n:$ $q=q_{m,n},$ $q_{1,0}=q_{m+1,n}$
and so on.  Moreover, it is convenient to introduce notations for the
differences (see figure below)
\[
    x=q-q_{-1,0}, \quad  y=q-q_{0,-1}, \quad  z=q-q_{-1,-1}.
\]
Obviously the following identity is valid
\be                                                         \label{xy}
    (T_m-1)y_{0,1} = (T_n-1)x_{1,0}.
\ee
The analog of the lattice (\ref{TL}) is given by 5-point difference equation
\be                                                         \label{xy5}
    (T_m-1)f(x) = (T_n-1)g(y)
\ee
of the discrete Toda lattice type \cite{Hir,Sur90,Sur95,AS}, and Legendre
transformation is of the form
\[
    Y_{0,1}=Q_{0,1}-Q=f(x), \quad X_{1,0}=Q_{1,0}-Q=g(y).
\]
It is not related to integrability as well as in continuous case.  The analog
of the relativistic lattices (\ref{RTL}) is more symmetric than its
continuous counterpart.  Let us consider the functionals of the form
\[
    {\cal L} = \sum_{m,n}(a(x) + b(y) + c(z)).
\]
Obviously, the rights of all terms (which play the roles of the kinetic and
potential energies and magnetic field) are absolutely equal, and therefore it
is more aesthetically to enumerate the variables $q$ by the points of the
rectilinear triangular lattice rather than square one.  Nevertheless we use
the standard grid which makes computations more easy.  Euler equation is of
the form
\be                                                         \label{xy7}
    (T_m-1)f(x) + (T_n-1)g(y) + (T_mT_n-1)h(z) = 0
\ee
where $f=a',$ $g=b',$ $h=c'$ and we assume that $f'g'h'\neq0$ in order
to exclude the case (\ref{xy5}).  Up to the author knowledge, the
equations of this type appeared for the first time in the papers
\cite{Sur96,Sur97a,Sur97b} by Suris, who also obtained them starting
from the relativistic Toda lattices.

The third difference can be expressed through $x,y$ in two different ways and
this circumstance allows to rewrite (\ref{xy7}) in two equivalent forms of
the momentum conservation law:
\[
         (T_m-1)(f(x)+h(z_{0,1})) + (T_n-1)(g(y)+h(z)) = 0  \LRA
\]\[
    \LRA (T_m-1)(f(x)+h(z)) + (T_n-1)(g(y)+h(z_{1,0})) = 0.
\]
As a result we can define a pair of the Legendre transformations
\be
    T_+:\ (X,Y_{-1,0})= T(x,y_{0,1}),\qquad
    T_-:\ (\ti X_{0,-1},\ti Y) = T(x_{1,0},y)               \label{Tpm}
\ee
where the mapping $T:(x,y)\to (X,Y)$ is given by formulae
\be                                                         \label{T}
    X = g(y)+h(x+y), \quad Y = -f(x) -h(x+y).
\ee                                                                                          \label{}

Notice, that the lattices (\ref{RTL}) are the continuous limit of the
equations (\ref{xy7}).  Indeed, let us consider the family of the functionals
\[
    {\cal L}_\eps = \sum_{m,n}(\eps a(x_{m,n}/\eps) + \eps b(y_{m,n})
                               - c(y_{m,n}) + c(z_{m,n})),
\]
and let $q_{m,n}=q_n(t),$ $t=m\eps.$  It is easy to see that passage to the
limit $\eps\to0$ brings to the functional
\[
    \int dt\sum_n (a(\dot q_n)+b_n+\dot q_nc'_n)
\]
which coincide with the one considered above, up to designations.  Legendre
transformations for the corresponding relativistic Toda lattice also are
obtained from (\ref{Tpm}) by the mentioned limit.

\begin{Def}
Equation (\ref{xy7}) is called integrable if the generalized Legendre
trans\-formations (\ref{Tpm}), (\ref{T}) are invertible (on the differences)
and map it into the equation of the same type.
\end{Def}

\begin{theorem}
Equation (\ref{xy7}) is integrable in the sense of the Definition 1 iff the
inverse of (\ref{T}) is of the form
\be                                                         \label{invT}
    x = G(Y)+H(X+Y), \quad  y = -F(X)-H(X+Y).
\ee
In this case the dual equation is
\be                                                         \label{XY7}
    (T_m-1)F(X) + (T_n-1)G(Y) + (T_mT_n-1)H(Z) = 0.
\ee
\end{theorem}
\proof
Obviously, it is sufficient to consider only transformation $T_+.$  Let
$T^{-1}_+$ be of the form stated above, then one can easily check that
identity (\ref{xy}) is equivalent to the equation (\ref{XY7}).  Conversely,
assume that $T_+$ maps (\ref{xy7}) into (\ref{XY7}).  Solving the formulae
$(X,Y_{-1,0})=T(x,y_{0,1})$ with respect to $x,y_{0,1}$ one obtains
\[
    x=\Phi(X,Y_{-1,0}), \quad y_{0,1}=\Psi(X,Y_{-1,0})
\]
and identity (\ref{xy}) yields equation
\[
   \Phi(X_{1,1},Y_{0,1})-\Phi(X_{1,0},Y)-\Psi(X_{1,0},Y)+\Psi(X,Y_{-1,0})=0
\]
which must be equivalent to (\ref{XY7}), that is
\[
   F(X_{1,0})-F(X) +G(Y_{0,1})-G(Y) +H(X_{1,1}+Y_{0,1}) -H(X+Y_{-1,0}) = 0.
\]
Let us consider $X_{1,1}$ as the function on the rest variables involved in
these equations, then
\[
    -{\pa X_{1,1}\over\pa Y_{0,1}}
    = {\pa\Phi\over\pa Y_{0,1}}\big/{\pa\Phi\over\pa X_{1,1}}
    = {G'(Y_{0,1})+H'(X_{1,1}+Y_{0,1})\over H'(X_{1,1}+Y_{0,1})}
\]
and hence $\Phi(X,Y)=\phi(G(Y)+H(X+Y)).$  Analogously the equality
\[
    -{\pa Y_{-1,0}\over\pa X}
    = {\pa\Psi\over\pa X}\big/{\pa\Psi\over\pa Y_{-1,0}}
    = {F'(X)+H'(X+Y_{-1,0})\over H'(X+Y_{-1,0})}
\]
yields $\Psi(X,Y)=\psi(F(X)+H(X+Y)).$  Finally, the relation
\[
    -{\pa X_{1,0}\over\pa Y}
    = {\pa(\Phi+\Psi)\over\pa Y}\big/{\pa(\Phi+\Psi)\over\pa X_{1,0}}
    = -{G'(Y)\over F'(X_{1,0})}
\]
proves $\Phi(X,Y)+\Psi(X,Y)=\chi(F(X)-G(Y)).$  Using these relations one
easily obtains $\phi(z)=\alpha z+\beta,$ $\psi(z)=-\alpha z+\gamma$ and
finish the proof (notice that functions $F,G,H$ from (\ref{XY7}) are defined
up to the linear transformation).
\qed
\medskip

The immediate corollary is that equation which is dual to the dual equation
coincides with the original one.  This allows to use composition
$T^{-1}_-T_+$ of the Legendre transformation for reproducing of the
solutions.

\section{Classification theorem}

So, in order to classify the integrable equations (\ref{xy7}) it is
sufficient to find all functions $f,g,h$ such that inverse of the
transformation (\ref{T}) is given by (\ref{invT}).  Hence the Jacobian
$\Delta=f'g'+g'h'+h'f'$ of the map (\ref{T}) must be nonzero and the
following identities must hold
\[
    x_{XY}+y_{XY}=0,\quad  x_{XY}=x_{XX},\quad  y_{XY}=y_{YY}.
\]
These three conditions are equivalent.  Indeed, the Jacobi matrix is
\[
    \left(\ba{cc} x_X & y_X \\ x_Y & y_Y \ea\right) =
    {1\over\Delta}
    \left(\ba{cc} -h' & f'+h' \\ -g'-h' & h' \ea\right)
\]
that is $x_X=-y_Y.$  Straightforward computation proves that functions
$f(x),$ $g(y),$ $h(x+y)$ must satisfy the equation
\[
    (g'+h'){f''\over f'} + (f'+h'){g''\over g'} = (f'+g'){h''\over h'}.
\]
Designations $f'=1/u,$ $g'=1/v,$ $h'=1/w$ rewrite it in more convenient form
\be                                                         \label{uvw}
    [v(y)+w(x+y)]u'(x) + [u(x)+w(x+y)]v'(y) = [u(x)+v(y)]w'(x+y).
\ee
The classification problem is reduced to solving of this functional equation.
Of course, the functions $u,v,w$ can be multiplied on the arbitrary constant
and linear transformation
\[
    \ti x=c(x-x_0), \quad \ti y=c(y-y_0) \LRA
    \ti q_{m,n}=c(q_{m,n}-mx_0-ny_0-\const)
\]
can be applied, as well as permutation of the coordinate axes.

At first let us consider some degenerate cases.  Assume that two of the
functions $u,v,w$ are constant, say $u$ and $v.$  Then (\ref{uvw}) yields
that either $w$ is constant as well or $u=-v$ and $w$ is arbitrary.  In the
first case the equation (\ref{xy7}) is linear and in the second it is of the
form
\[
   \alpha(q_{1,0}+q_{-1,0}-q_{0,1}-q_{0,-1})+h(q_{1,1}-q)-h(q-q_{-1,-1}) = 0
\]
and admits one integration
\[
   \alpha(q_{m+1,n}-q_{m,n+1})+h(q_{m+1,n+1}-q_{m,n}) = c_{m-n}.
\]
Other degenerate case corresponds to the case of zero Jacobian, that is
$u+v+w=0$ (this implies that all three functions are linear).  In this case
one can prove that equation (\ref{xy7}) can be reduced to the equation on the
variable $p=y_{0,1}/x.$  Further on we will not consider these degenerate
cases.

\begin{lemma}
Functions $u,v,w$ satisfy equations
\be                                                         \label{uvw'}
    (u')^2= \delta u^2+2\alpha u+\eps,\quad
    (v')^2= \delta v^2+2\beta  v+\eps,\quad
    (w')^2= \delta w^2+2\gamma w+\eps.
\ee
\end{lemma}
\proof
At first prove that functions $u(x)$ and $v(y)$ satisfy equation
\be                                                         \label{uv}
    (u''-v'')(u+v)-(u')^2+(v')^2=k(u-v), \quad  k=\const.
\ee
Let us eliminate $w$ from (\ref{uvw}).  Applying the operator $\pa_x-\pa_y$
one obtains the linear system on $w,w':$
\[
    \left(\ba{cc}
      u'+v'  & -u-v \\
     u''-v'' & v'-u'
    \ea\right) \left(\ba{c} w+u \\ w'  \ea\right) =
          (u-v)\left(\ba{c}  u' \\ u'' \ea\right).
\]
Its determinant $\Delta$ is exactly the left hand side of the equation
(\ref{uv}).  If it is identically zero then (\ref{uv}) is proved, otherwise
one finds
\[
    w+u= {u-v\over\Delta}(uu''-(u')^2+u''v+u'v'),\quad
     w'= {u-v\over\Delta}(u''v'+u'v'')
\]
and consequently $((u-v)/\Delta)_y(uu''-(u')^2+u''v+u'v')=0.$  Assume that
the expression in the second bracket vanishes.  If $u'\not\equiv0$ then
$v'=-u''v/u'+u'-uu''/u',$ $v''=-u''v'/u',$ but then, as one easily checks,
$\Delta=0.$  If $u'=0$ then $w+u=0$ that is we come to the degenerate
solution excluded above.  Therefore $((u-v)/\Delta)_y=0.$  Due to the
symmetry between $u$ and $v$ one can prove analogously $((u-v)/\Delta)_x=0$
and obtain (\ref{uv}).

Further on, rewriting (\ref{uv}) in the form
\[
   \left({u'\over u+v}\right)_x = {v''+k\over u+v}-{(v')^2+2kv\over(u+v)^2}
\]
multiplying by $u'/(u+v)$ and integrating with respect to $x$ yield
\[
    (u')^2 =    \delta(y)(u+v)^2 - 2(v''+k)(u+v)+(v')^2+2kv.
\]
Replacing $u$ and $v$ one obtains
\[
    (v')^2 = \ti\delta(x)(u+v)^2 - 2(u''+k)(u+v)+(u')^2+2ku.
\]
Subtracting one equation from the other one and using (\ref{uv}) one obtains
$\ti\delta=\delta=\const.$  Summing and dividing by $u+v$ give
$u''+v''=\delta(u+v)-k.$  Separation of variables yields $(u')^2=\delta
u^2+2\alpha u+\eps,$ $(v')^2=\delta v^2+2\beta v+\ti\eps,$ where
$\alpha+\beta=-k,$ and substitution into (\ref{uv}) proves $\eps=\ti\eps.$
The last of the equations (\ref{uvw'}) is obtained in virtue of symmetry of
the coordinate axes.
\qed
\medskip

It is clear that solutions of the equations (\ref{uvw'}) must satisfy also
some additional relations.  However their analysis is not in principle
difficult, and direct examination of all solutions brings to the following
list.

\begin{theorem}
The equations (\ref{xy7}) integrable in the sense of Definition 1, are
exhausted, up to changes $\ti q_{m,n}=c(q_{m,n}-mx_0-ny_0)$ and permutations
of coordinate axes, by the following sets of the functions $f,g,h.$  In
formulae (A), (B), (C) parameters are constrained by relation
$\la+\mu+\nu=0,$ and in (I) by relation $\la\mu\nu=-1.$
\[\ba{lllll}
 (A) && f={\mu\over x}, &  g={\nu\over y}, &  h={\la\over z}, \\
 (B) && f=\mu\coth x,   &  g=\nu\coth y,   &  h=\la\coth z,   \\
 (C) && f=\12\log{x+\mu\over x-\mu}, &
        g=\12\log{y+\nu\over y-\nu}, &
        h=\12\log{z+\la\over z-\la},  \\
 (D) && f=\log x,          &  g=\log y,     &  h=\log(1-1/z),     \\
 (E) && f=-e^x-1,          &  g=e^{-y},     &  h={1\over 1+e^z},  \\
 (F) && f=\log(e^x-1),     & g=\log(e^y-1), &  h=-\log(e^z-1),    \\
 (G) && f=-\log(e^{-x}-1), & g=\log(e^y-1), &  h=-z,              \\
 (H) && f=\log(\la^{-1}(e^x+1)),   &
        g=\log(e^{-y}-1),          &
        h=\log{e^z+\la\over e^z+1}, \\
 (I) && f=\log{\mu e^x+1\over e^x+\mu}, &
        g=\log{\nu e^y+1\over e^y+\nu}, &
        h=\log{\la e^z+1\over e^z+\la}.
\ea\]
Legendre transformations (\ref{Tpm}), (\ref{T}) link together the equations
corresponding to solutions (B) and (C), (D) and (E), (F) and (G), while
equations corresponding to solutions (A), (H) and (I) are self-dual.
\qed
\end{theorem}

Notice, that cases (A) and (B) are connected by point transformation
$q=\exp(2\ti q).$  It is explained by fact that the Lagrangian $\sum(\mu\log
x+\nu\log y+\la\log z)$ of the equation (\ref{xy7}), (A) is invariant under
the dilations $q\to Cq$ as well as under the shifts $q\to q+C.$  From the
other hand, inversions $q\to q/(1-Cq)$ preserves Lagrangian as well but the
change $q=1/\ti q$ which maps this group into the shift group does not bring
to new equation.

Apparently, the cases (E), (F), (G), (H) and (A), (B) for $\la=0$ appeared in
the papers \cite{Sur95,Sur96,Sur97a,Sur97b} by Suris for the first time.  The
rest cases are new, up to the author knowledge.

\section{Higher symmetries}

Study of the properties of the presented equations goes out of the scope of
this paper.  Of course, one has to prove that the Definition 1 brings to
equations which are integrable in more habitual sense, that is satisfying the
Painlev\'e test or possessing the higher symmetries and conservation laws.
Now we restrict ourself by presenting of the zero curvature representation
and some higher symmetries for the most simple equation (\ref{xy7}), (A),
that is
\be                                                         \label{Aq}
    {\mu\over q_{1,0}-q}+{\mu\over q_{-1,0}-q}+
    {\nu\over q_{0,1}-q}+{\nu\over q_{0,-1}-q}+
    {\la\over q_{1,1}-q}+{\la\over q_{-1,-1}-q}=0,
\ee
where $\la+\mu+\nu=0.$  Notice, that all formulae in this section remain
valid also for vanishing $\la$ when the equation (\ref{Aq}) degenerates into
equation of the form (\ref{xy5}).

(Continuous) symmetry of the discrete equation $E=0$ is a vector field
$q_t=\Phi$ on the lattice which preserves this equation:
$\pa_t(E)=0|_{E=0}.$  At first let us consider, for completeness, the classic
symmetries $q_t=1,$ $q_t=q,$ $q_t=q^2$ corresponding to the group of
linear-fractional transformations.  Since these symmetries are variational,
hence in virtue of the Noether theorem multiplying of (\ref{Aq}) by their
right hand sides yields some conservation laws.  Indeed, the shift symmetry
generates the momentum conservation law (\ref{xy7}), and dilations and
inversions give respectively the conservation laws
\[
      (T_m-1){\mu q\over q-q_{-1,0}}
    + (T_n-1){\nu q\over q-q_{0,-1}}
    + (T_mT_n-1){\la q\over q-q_{-1,-1}} = 0,
\]\[
      (T_m-1){\mu qq_{-1,0}\over q-q_{-1,0}}
    + (T_n-1){\nu qq_{0,-1}\over q-q_{0,-1}}
    + (T_mT_n-1){\la qq_{-1,-1}\over q-q_{-1,-1}} = 0.
\]

The structure of the higher symmetries of the equation (\ref{Aq}) is rather
original, as one can infer out of the simplest representatives, and it would
be very interesting to obtain the complete description of the symmetry
algebra.

\begin{theorem}
The following formulae define the symmetries of the equation (\ref{Aq}):
\[
    {1\over q_\xi}   = {\mu\over q_{-1,0}-q} +{\nu\over q_{0,1}-q}
                      +{\la\over q_{-1,-1}-q},        \quad
    {1\over q_\eta}  = {\mu\over q_{1,0}-q}+{\nu\over q_{0,-1}-q}
                      +{\la\over q_{-1,-1}-q},
\]\[
    {1\over q_\zeta} = {\mu\over q_{1,0}-q}+{\nu\over q_{0,1}-q}
                      +{\la\over q_{1,1}-q}.
\]
Vector fields $\pa_\xi,\pa_\eta,\pa_\zeta$ commute in virtue of the equation
(\ref{Aq}). \qed
\end{theorem}

The proof can be obtained by straightforward, although tedious computation.
It is sufficient to check the determining equation $\pa_t(E)=0|_{E=0}$ only
for one symmetry, because two others are obtained by cyclic permutation of
the axes $x,y,z.$

The zero curvature representation for the equation (\ref{Aq}) is defined as
compati\-bility condition for the auxiliary linear problems
$\Psi_{1,0}=A\Psi$ and $\Psi_{1,1}=B\Psi_{1,0}.$  In order to retain the
equal rights of the directions on the lattice, let us consider also the
equation $\Psi=C\Psi_{1,1}$ and obtain
\be                                                         \label{BAC}
    BA=A_{0,1}B_{-1,0},\quad AC=C_{1,0}A_{1,1},\quad CB=B_{-1,-1}C_{0,-1}.
\ee
Consistency with the problem of the form $\Psi_t=U\Psi$ corresponding to the
continuous symmetry, brings to the equations
\be                                                         \label{U}
    A_t=U_{1,0}A-AU,\quad B_t=U_{1,1}B-BU_{1,0},\quad C_t=UC-CU_{1,1}.
\ee
\begin{figure}[t]
\setlength{\unitlength}{0.07em}
\begin{picture}(290,240)(-230,-100)
 \put(-115,0){\vector(1,0){240}}
 \put(0,-115){\vector(0,1){240}}
 \put(-100,-100){\line(1,1){200}}
 \multiput(-100,-100)(200,100){2}{\line(0,1){100}}
 \multiput(-100,-100)(100,200){2}{\line(1,0){100}}
 \multiput(-100,0)(100,-100){2}{\line(1,1){100}}
 \put(130,-20){$T_m$} \put(-25,120){$T_n$}  \large
 \put(-80,-35){$\Psi$}       \put(-50,-30){\vector(1,0){60}}
 \put( 20,-35){$\Psi_{1,0}$} \put( 35,-10){\vector(0,1){60}}
 \put( 20, 70){$\Psi_{1,1}$} \put( 10, 50){\vector(-1,-1){60}} \Large
 \put(-35,-65){$A$}
 \put( 60, 25){$B$}
 \put(-40, 30){$C$}
\end{picture}\normalsize
\end{figure}

Matrices $A,B$ and $C$ are of the form
\bean
  && A= (k-\mu)I - \nu P(q,q_{0,-1}) - \la P(q,q_{-1,-1}), \\
  && B= kI       - \mu P(q,q_{1,0})  - \la P(q,q_{1,1}),   \\
  && C= (k+\nu)I - \mu P(q,q_{-1,0}) - \nu P(q,q_{0,1}),
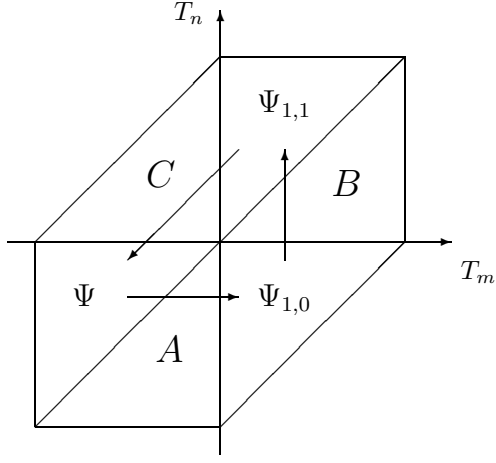
\eean
where $k$ is spectral parameter, $I$ is unit matrix and projector $P$ is
given by formula
\[
    P(u,v)={1\over u-v}\left(\ba{cc} -v & -uv \\ 1 & u \ea\right)
\]
(the coefficients on $I$ can be chosen more symmetric by shift $k\to
k+(\mu-\nu)/3).$)  On the picture above each of the matrices $A,B,C$ depends
on the vertices of the triangle in which it is placed (the center of the
hexagon corresponds to the variable $q$).  Notice also that successive
application of the operators $A,B,C$ maps the wave function into itself up to
the scalar factor, since $CBA=k(k-\mu)(k+\nu)I.$  The check of the
equivalence of (\ref{Aq}) and equations (\ref{BAC}) is not difficult when
using the properties
\[
    P(u,v)P(p,q)={(u-q)(p-v)\over(u-v)(p-q)}P(p,v), \quad P(u,v)+P(v,u)=I.
\]

The representations (\ref{U}) for the symmetry $\pa_t=\pa_\zeta$ are given by
the matrix $U=k^{-1}(\ti U-{1\over2}I),$ where $\ti U$ is the projector
\[
  \ti U = {1\over\alpha+q_{-1,0}\beta}
            \left(\ba{cc}
              \alpha & q_{-1,0}\alpha \\
              \beta  & q_{-1,0}\beta
            \ea\right),
\]
$\alpha = \la qq_{-1,-1} + \mu q_{-1,-1}q_{0,-1} + \nu q_{0,-1}q,$
$\beta = \la q_{0,-1} + \mu q + \nu q_{-1,-1}.$

It should be stressed that the symmetries presented in Theorem 3 are not
integrable equations by themselves and it makes sense to consider them only
together with the constraint (\ref{Aq}).  In other words, they are 1+1 rather
than 1+2-dimensional equations.  Indeed, denoting $u_m=q_{m,n},$
$v_m=q_{m,n-1}$ and using equation (\ref{Aq}) one can rewrite the symmetries
$\pa_\eta,$ $\pa_\zeta$ as the one-dimensional lattices
\be                                                         \label{eta}
    {1\over u_\eta} = {\mu\over u_1-u} + {\nu\over v-u}
                      +{\la\over v_{-1}-u},  \quad
   -{1\over v_\eta} = {\mu\over v_{-1}-v} + {\nu\over u-v}
                      +{\la\over u_1-v},
\ee
\be                                                         \label{zeta}
   -{1\over u_\zeta}= {\mu\over u_{-1}-u} + {\nu\over v-u}
                     +{\la\over v_{-1}-u},  \quad
    {1\over v_\zeta}= {\mu\over v_1-v} + {\nu\over u-v}
                     +{\la\over u_1-v}
\ee
(in order to write down $\pa_\xi$ in analogous form one should choose
dynamical variables along the axis $y$ or $z.$)

Remarkable fact is that these lattices are equivalent to one of the
relativistic lattices (\ref{RTL}), namely to the lattice, corresponding to
the isotropic Heisenberg model.

\begin{theorem}
Elimination of the variables $v_m$ from the equations (\ref{eta}) brings to
the lattice
\[
    u_{\eta\eta} = u^2_\eta\left(
                     {\mu u_{1,\eta}\over (u_1-u)^2}
                   - {\mu u_{-1,\eta}\over (u-u_{-1})^2}
                   - {1\over u_1-u} + {1\over u-u_{-1}} \right).
\]
Vector fields defined by equations (\ref{eta}) and (\ref{zeta}) commute and
variables $u=u_m,$ $v=v_m$ satisfy the system (now we assume that $\mu\neq0$)
\[
    u_{\eta\zeta} = {2u_\eta u_\zeta\over u-v} + {u_\eta\over\mu}
     - {u_\zeta\over\mu}\sqrt{1-4\la\mu{u_\eta v_\eta\over(u-v)^2}}
\]\[
    v_{\eta\zeta} = {2v_\eta v_\zeta\over v-u} - {v_\eta\over\mu}
     + {v_\zeta\over\mu}\sqrt{1-4\la\mu{u_\eta v_\eta\over(u-v)^2}}
\]
in virtue of these equations.
\qed
\end{theorem}

So, the relation between the continuous equations (\ref{RTL}) and their
discrete analogues (\ref{xy7}) is more profound than it seemed at first
sight.  Remind, that the relation between the lattices (\ref{RTL}) and
equations of the form (\ref{xy5}) has been already established in paper
\cite{AS}.  More precisely, it was demonstrated there that the composition of
the Legendre transformations $T^{-1}_-T_+$ for relativistic Toda lattices is
equivalent to the shift in Toda lattices (\ref{TL}) (however, the analogy
with discrete equations is slightly failed at this point, since the
composition $T^{-1}_-T_+$ of the transformations (\ref{Tpm}) cannot be
rewritten as equation of the form (\ref{xy5})) and the nonlinear
superposition principle for the lattices of these two types brings to the
equations (\ref{xy5}).  The example considered above demonstrates that the
equations (\ref{xy7}) can be interpreted as nonlinear superposition principle
as well, but for the pair of relativistic lattices.

In conclusion of this section notice that, although the equation (\ref{Aq})
was not subjected to the Painlev\'e test, but the simplest examples of
2-periodic reductions (several variants are possible) demonstrate that it is
Painlev\'e integr\-able even at $\la+\mu+\nu\neq 0.$  Of course in this case
the above zero curvature representations and symmetries are lacking.
Possibly, this situation is analogous to the case of difference Hirota-Miwa
and KdV equations which were verified by the Painlev\'e test in the paper
\cite{RGS}.

\section{Concluding remarks}

The author's main purpose was to demonstrate that condition of invariance
with respect to the Legendre transformations allows effectively isolate the
class of integrable equations.  The obtained examples of difference equations
at the triangular lattice provide the discrete analogues of relativistic Toda
lattices and prompt expectation that the method of Legendre transformations
suggested in \cite{AS} is not just a trick and its scope is rather wide.  The
open problem is further generalizations on the new classes of equations, in
particular multidimensional ones.

As in continuous case the weakest point of the method is the conjecture about
the shift invariance of the Lagrangian.  The study of the relativistic Toda
lattices \cite{AS} implies that the difference analog of the Landau-Lifshitz
equation must exist which is characterized by absence of the classic
symmetries.

>From the other hand, it is not difficult to find some multifield
generalizations, although their classification is far from completeness, as
well as in continuous case.  Now we present only the vector analog of the
discrete Heisenberg equation (\ref{Aq}):
\[
    \mu(T_m-1){x\over|x|^2} + \nu(T_n-1){y\over|y|^2}
             = (\mu+\nu)(T_mT_n-1){z\over|z|^2}
\]
where $q\in\Real^d$ and $|q|$ denotes Euclidean norm.  Legendre
transformation $T_+$ is of the form
\[
 X=\nu{y_{0,1}\over|y_{0,1}|^2}-(\mu+\nu){x+y_{0,1}\over|x+y_{0,1}|^2}, \quad
 Y_{-1,0}= -\mu{x\over|x|^2}+(\mu+\nu){x+y_{0,1}\over|x+y_{0,1}|^2}
\]
and, as in the scalar case, maps equation into itself.  Apparently this
example admits generalizations for arbitrary Jordan triple systems with
invertible elements.


\end{document}